\documentclass[aps,pra,twocolumn,groupedaddress]{revtex4}

\usepackage{graphicx}

\newcommand{\rs}{\rm \scriptscriptstyle}
\newcommand{\bs}{\bf \scriptscriptstyle}
\begin{document}

\title{Phase separation of atomic Bose-Fermi mixtures in an optical lattice}

\author{H.P.\ B\"uchler}
\author{G.\ Blatter}
\affiliation{Theoretische Physik, ETH-H\"onggerberg, CH-8093
Z\"urich, Switzerland}

\date{\today}

\begin{abstract}
  We study a two-dimensional atomic mixture of bosons and fermions
  cooled into their quantum degenerate states and subject to an
  optical lattice. The optical lattice provides van Hove
  singularities in the fermionic density of states. We find
  that these van Hove singularities produce new and interesting
  features for the transition towards phase separation:
  an arbitrary weak interaction between the bosons and the
  fermions is sufficient to drive the phase separation at low
  temperatures. The phase separated state turns stable
  for attractive and repulsive interaction between the bosons
  and fermions and can be cast into the standard form of a
  `liquid--gas' transition.
\end{abstract}


\maketitle

\section{Introduction}

Atomic mixtures of  bosons and fermions represent a new laboratory
system for the study of quantum degenerate matter. A major
experimental breakthrough is the sympathetic cooling of atomic
fermions into their quantum degenerate state
\cite{truscott01,schreck01,hadzibabic02,roati02} allowing for the
realization of new thermodynamic phase transitions. The main focus
is on the superfluid transition of atomic fermions
\cite{houbiers97,heiselberg00,hofstetter02,viverit02}, and the
appearance of complex quantum phases in the presence of an optical
lattice, such as supersolids \cite{buechler03.2} and nontrivial
Mott insulating phases \cite{lewenstein03}. In addition, a
Bose-Fermi mixture cooled into its quantum degenerate state shows
a strong tendency towards phase separation and the demixing of the
bosons and fermions \cite{molmer98,amoruso98}. While this phase
separation itself exhibits many fascinating phenomena
\cite{nygaard99,modugno02,viverit00,roth02}, it also represents a
major drawback of Bose-Fermi mixtures as it restricts the allowed
parameter range for the observation of complex quantum phases, see
e.g., the competition between phase separation and a supersolid
phase as discussed in Ref.\ \onlinecite{buechler03.2}.  The
analysis of phase separation in Bose-Fermi mixtures then
contributes to our basic understanding of atomic gases as needed
in future studies of other novel quantum phases.

The main mechanism driving the phase separation is a small
perturbation of the bosonic density $\delta n_{\rs B}$ inducing a
modulation of the fermionic density $\delta n_{\rs F}\sim - U_{\rs
FB} N(\epsilon_{\rs F}) \delta n_{\rs B}$. Here, $N(\epsilon_{\rs
F})$ denotes the density of states at the Fermi energy
$\epsilon_{\rs F}$ and $U_{\rs FB}$ is the strength of the
interaction between the bosons and the fermions. This fermionic
distortion acts back onto the bosons and produces a shift of the
bosonic energy $-U_{\rs FB}^{2} N(\epsilon_{\rs F}) \delta n_{\rs
B}^{2}/2$, thereby inducing an attraction between the bosons with
strength $U_{\rs FB}^{2} N(\epsilon_{\rs F})$. Phase separation
appears if the induced attraction is of the same order as the
intrinsic repulsion $U_{\rs B}$ between the bosons.

Previous work on phase separation in Bose-Fermi mixtures
concentrated on the $T=0$ limit. The phase separation is
triggered by changing the interaction $U_{\rs FB}$ and results in
minimal energy configurations involving separated states with two
pure phases, two differently mixed phases, or one pure phase
coexisting with one mixed phase. Assuming a repulsive interaction 
between the fermions and the bosons, the phase separation of 
Bose-Fermi mixtures in a 3D harmonic trap results in an 
increased bosonic density accumulated in the center of the
trap and surrounded by a fermionic shell 
\cite{molmer98,nygaard99,amoruso98}. A thermodynamic analysis 
of the homogeneous situation \cite{viverit00} has shown
that the criterion for phase separation described above is 
preempted by a first-order scenario resulting in a separation 
involving a mixed and a pure phase. On the other hand, an 
attractive interaction between the bosons and fermions 
triggers a system collapse above a critical boson number
\cite{roth02}; such a collapse has been recently observed in an
experiment by Modugno {\it et al.} \cite{modugno02}.

Here, we concentrate on the behavior of phase separation in a
situation, where the fermionic density of states is strongly
modified by the presence of an optical lattice. Of special
interest is the two-dimensional (2D) situation where the presence
of saddle points in the fermionic dispersion relation
$\epsilon_{\rs F}({\bf k})$ provides logarithmic van Hove
singularities in the density of states $N(\epsilon)\sim N_{0}
\ln|2W_{\rs F}/(\epsilon - \epsilon_{0})|$, with $W_{\rs F}$ the
band width and $\epsilon_0$ the position of the van Hove
singularity. A finite temperature $T$ cuts this singularity 
and the system response becomes strongly temperature
dependent, resulting in a smooth transition with decreasing
temperature into a phase-separated state with two mixed phases.
Below, we present a detailed analysis of this finite temperature
phase transition. We find that the van Hove singularities strongly
enhance the tendency towards phase separation, with an arbitrary 
weak interaction between the bosons and the fermions already
sufficient to drive the transition at low temperatures. The
transition is of the `liquid--gas' transition type and the phase
separated state is stable for attractive as well as for
repulsive interaction between the bosons and fermions. 
Concentrating on the weak coupling limit, we derive the 
$n_{\rs F}$-$T$ phase diagram involving two mixed Bose-Fermi 
gases with slightly different fermion density $\delta n_{\rs F}$. 
Here, we focus on phase separation and ignore other possible 
instabilities in the system, e.g.,
BCS-superconductivity or supersolid formation. Note that these
competing phases also exhibit a strong increase of the critical
temperature due to the van Hove singularity, a phenomenon well
studied in BCS-superconductivity \cite{hirsch86}. Furthermore, 
our weak coupling analysis with $U_{\rs FB}^{2} N_{0} \ll 
U_{\rs B}$ excludes the appearance of a competing first-order 
phase transition of the type discussed in Ref.\ 
\onlinecite{viverit00}.

The Hamiltonian of an interacting Bose-Fermi mixture subject to an
optical lattice is presented in Sec.~\ref{bosefermimixtures}. We
make use of the tight-binding approximation to cast the
Hamiltonian in a form describing  a Bose-Fermi mixture on a
lattice. We then integrate out the fermions and arrive at an
effective boson Hamiltonian. The interaction becomes strongly
temperature dependent and  the effective boson Hamiltonian exhibits
an instability towards phase separation at a critical temperature
$T_{\rs PS}$. This instability is further examined in
Sec.~\ref{phaseseparation} and the phase separated state is
studied within the Thomas-Fermi description; the results are
summarized in a phase diagram exhibiting all the properties
of a standard liquid-gas transition.
Finally, we study the impact of this transition on an atomic
Bose-Fermi mixture in a finite trap in Sec.~\ref{conclusions}.

\section{Bose-Fermi mixtures \label{bosefermimixtures}}

We start with interacting bosons and fermions in two dimensions.
Such a 2D setup is achieved via application of a strong confining
potential in the transverse direction. The Hamiltonian for interacting
bosons and fermions subject to an optical lattice takes the form $H =
H_{\rm \scriptscriptstyle B} + H_{\rm \scriptscriptstyle F} +
H_{\rm \scriptscriptstyle int}$ with
\begin{eqnarray}
    H_{\rm \scriptscriptstyle B}&=&\int d{\bf x} \:\:
     \psi^{+}_{\rm \scriptscriptstyle  B}
     \left( - \frac{\hbar^{2}}{2m_{\rm \scriptscriptstyle B}} \Delta +
      V_{\rm \scriptscriptstyle B}({\bf x}) \right)
      \psi_{\rm \scriptscriptstyle  B},  \nonumber\\
 H_{\rm \scriptscriptstyle F} &=&\int d{\bf x} \:\:
     \psi^{+}_{\rm \scriptscriptstyle  F}
     \left( - \frac{\hbar^{2}}{2m_{\rm \scriptscriptstyle F}} \triangle +
      V_{\rm \scriptscriptstyle F}({\bf x}) \right)
      \psi_{\rm \scriptscriptstyle  F},  \label{startinghamiltonian3}\\
H_{\rm \scriptscriptstyle int}& = & \int d{\bf x} \left(g_{\rm
\scriptscriptstyle FB}  \psi^{+}_{\rm \scriptscriptstyle B}
      \psi_{\rm \scriptscriptstyle  B} \psi^{+}_{\rm \scriptscriptstyle  F}
      \psi_{\rm \scriptscriptstyle  F} + \frac{1}{2}g_{\rm \scriptscriptstyle B}
      \psi^{+}_{\rm \scriptscriptstyle  B}\psi^{+}_{\rm \scriptscriptstyle  B}
      \psi_{\rm \scriptscriptstyle  B}
      \psi_{\rm \scriptscriptstyle  B} \right). \nonumber
\end{eqnarray}
Here, $\psi_{\rs F}$ and $\psi_{\rs B}$ denote the fermionic and
bosonic field operators, while the interaction between the
particles is taken into account within the pseudopotential
approximation. We assume a repulsive interaction $g_{\rm
\scriptscriptstyle B}= 4 \pi a_{\rs B} \hbar^{2}/m_{\rs B}$
between the bosons with the scattering length $a_{\rs B}>0$, while
the coupling $g_{\rs FB}=2 \pi a_{\rs FB} \hbar^{2}/\mu$ accounts
for the interaction between the fermions and the bosons with $\mu$
the relative mass and $a_{\rs FB}$ the boson-fermion scattering
length of either sign. Furthermore, we restrict the analysis to
spinless fermions; such a spinless fermionic atomic gas is
naturally achieved in an experiment via spin polarization. Then,
the $s$-wave scattering length of the fermion-fermion interaction
vanishes, while $p$-wave scattering is suppressed at low energies
and is neglected in the following analysis. The optical lattice
with wave length $\lambda$ provides an $a=\lambda/2$-periodic
potential for the bosons ($\alpha={\rm B}$) and fermions
($\alpha={\rm F}$) with $V_{\alpha}({\bf x}) =V_{\alpha}\sin^{2}(
\pi x /a) +V_{\alpha} \sin^{2}( \pi y / a)+V^t_\alpha$, while
the trapping potential $V^t_\alpha$ accounts for the strong
transverse confining $m_{\alpha} \omega_{\alpha\perp}^{2} z^{2}/2$
establishing a two dimensional setup, and for the weak
longitudinal trapping $m_{\alpha} \omega_{\alpha \parallel}^{2}
(x^{2}+y^{2})/2$.

\subsection{Hamiltonian within tight-binding approximation}

In the following, we focus on strong optical lattices $V_{\alpha}>
E_{\alpha}= 2 \hbar^{2}\pi^{2}/\lambda^{2} m_{\alpha}$ and
weak interactions. Then, in analogy to the mapping onto the
Bose-Hubbard model \cite{jaksch98}, we transform the Hamiltonian
(\ref{startinghamiltonian3}) for the Bose-Fermi mixture to a simplified
Hamiltonian of the tight-binding form. We restrict the analysis to
the lowest Bloch band and introduce the Bloch wave functions
$v_{k}$ (fermions) and $w_{k}$ (bosons) of the corresponding
single particle problem in the 2D periodic potential.  In turn,
the Bloch wave functions $w_{{\bf k}}({\bf x})$ and $v_{{\bf
k}}({\bf x})$ are related to the Wannier functions
$\widetilde{w}({\bf x-R})$ and $\widetilde{v}({\bf x-R})$
according to
\begin{eqnarray}
   \widetilde{w}({\bf x-R}) &=& \frac{1}{N}
   \sum_{{\bf k}\in K} w_{{\bf k}}({\bf
   x})\exp\left(-i {\bf R}{\bf k}\right),\\
   \widetilde{v}({\bf x-R})
   &=& \frac{1}{N}\sum_{{\bf k}\in K} v_{{\bf k}}({\bf x})
   \exp\left(-i {\bf R}
   {\bf k}\right),
\end{eqnarray}
with ${\bf R}$ a lattice vector and $K$ the first Brillouin zone
of the reciprocal lattice. Here, we have introduced the
quantization volume $V=N a^{2}$ with $N$ the number of unit cells.
In the following, $n_{\rs F,B}$ denote the number of particles per
unit cell. We express the bosonic and fermionic field operators
$\psi_{\rm \scriptscriptstyle  F, B} $ in terms of the Bloch wave
functions $w_{{\bf k}}$ and $v_{{\bf k}}$, or equivalently, in
terms of the Wannier functions $\widetilde{w}$ and $\widetilde{v}$
\begin{eqnarray}
    \psi_{\rm \scriptscriptstyle  B}^{+}({\bf x})
    &=&\frac{1}{\sqrt{N}} \sum_{{\bf k}\in K}
    w_{{\bf k}}({\bf x}) b^{+}_{{\bf k}}
    = \sum_{{\bf R}} \widetilde{w}({\bf
    x-R})\:b^{+}_{\bs R},
    \\ \psi_{\rm \scriptscriptstyle  F}^{+}({\bf x})
    &=& \frac{1}{\sqrt{N}}\sum_{{\bf k}\in K}
    v_{{\bf k}}({\bf x}) c^{+}_{{\bf k}}
    = \sum_{{\bf R}} \widetilde{v}({\bf x-R}) \: c^{+}_{\bs R}.
  \label{fieldoperators}
\end{eqnarray}
The bosonic creation operator $b^{+}_{{\bf k}}$ of the
Bloch wave state $w_{\bf k}$ is the  Fourier transform of the
bosonic creation operator $b^{+}_{{\bs R}}$ for the Wannier state
$\widetilde{w}({\bf x-R})$ with
\begin{equation}
    b^{+}_{{\bs R}}
    = \frac{1}{\sqrt{N}}\sum_{k \in K} \exp(i {\bf kR})\, b^{+}_{{\bf k}},
\end{equation}
and analogously for the fermionic creation operators $c^{+}_{{\bf
k}}$ and $c^{+}_{{\bs R}}$. In our subsequent analysis it is
convenient to use the operators $b^{+}_{{\bf k}}$ and $c^{+}_{{\bf
k}}$. Inserting the expansion (\ref{fieldoperators}) into the
Hamiltonian (\ref{startinghamiltonian3}) and restricting the
analysis to on-site interactions, the Hamiltonian of the
Bose-Fermi mixture (\ref{startinghamiltonian3}) reduces to
\begin{eqnarray}
    H &= &\sum_{{\bf k} \in K} \epsilon_{\scriptscriptstyle \rm B}({\bf k})
    b^{+}_{{\bf k}} b^{}_{{\bf k}}
    +\frac{U_{\scriptscriptstyle \rm B}}{2N} \sum_{\{{\bf k,k',q,q'}\}}
    b^{+}_{{\bf k}}  b^{+}_{{\bf q}}b^{}_{{\bf k}'} b^{}_{{\bf q}'}
    \label{tbhamiltonian}\\
    & &+ \sum_{{\bf q} \in K} \epsilon_{\scriptscriptstyle \rm
    F}({\bf q}) c^{+}_{{\bf q}}c^{}_{{\bf q}}
    + \frac{U_{\scriptscriptstyle \rm FB}}{N} \sum_{\{{\bf k,k',q,q'}\}}
    b^{+}_{ {\bf k}} b^{}_{{\bf k}'} c^{+}_{{\bf q}}c^{}_{{\bf
    q}'} \nonumber.
\end{eqnarray}
The summation $\{ {\bf k,k',q,q'}\}$ is restricted to ${\bf
k,k',q,q'} \in K$ and the momentum conservation ${\bf k}- {\bf
k}'+{\bf q}-{\bf q}'={\bf K}_{m}$ with ${\bf K}_{m}$ a reciprocal
lattice vector; a scattering event involving such a vector
${\bf K}_{m}$  is an Umklapp process. The interaction parameters
involve the Wannier functions $\widetilde{w}$ and $\widetilde{v}$
and take the form
\begin{eqnarray}
    U_{\scriptscriptstyle
    \rm B}&= &g_{\rs B}\int d{\bf x}|\widetilde{w}({\bf x})|^{4},
    \label{UB}\\
    U_{\scriptscriptstyle \rm FB}&=& g_{\rs FB}\int d{\bf x}
    |\widetilde{w}({\bf x})|^{2}|\widetilde{v}({\bf x})|^{2},
   \label{UFB}
\end{eqnarray}
while $\epsilon_{\rs F}({\bf k})$  and $\epsilon_{\rs B}({\bf k})$
denote the lowest energy bands for the fermions and bosons,
respectively.

In a two dimensional system, the band structure $\epsilon_{\rs F}
({\bf k})$ exhibits saddle points at wave vectors ${\bf k}_{0i}$
and energies $\epsilon_{0i}=\epsilon_{\rs F}({\bf k}_{0i})$.
Lattice symmetries naturally produce several saddle points at the
same energy $\epsilon_{0}$; we denote their number by $z$. Close
to such a saddle point, the dispersion relation $\epsilon_{\rs
F}({\bf k})$ is quadratic, and can be expanded along the principal
axes (${\bf k} = {\bf k}_{0}+{\bf k}_{\perp}+{\bf k}_{\parallel}$)
\begin{equation}
 \epsilon_{\rs F}({\bf k}) \sim \epsilon_{0}
  + \frac{\hbar^2}{2 m_{\perp}} {\bf k}_{\perp}^{2}
  -\frac{\hbar^2}{2 m_{\parallel}}{\bf k}_{\parallel}^{2}.
\end{equation}
The saddle point gives rise to van Hove singularities in the
density of states (per site) for $\epsilon \rightarrow 
\epsilon_{0}$,
\begin{equation}
N\left(\epsilon\right) \sim N_{0} \ln \left|\frac{ 2 W_{\rs
F}}{\epsilon-\epsilon_{0}} \right| \label{densityofstates}
\end{equation}
with the band width $W_{\rs F}$, and the prefactor $N_{0}= z
(m_{\perp} m_{\parallel})^{1/2} (a^{2}/2 \pi^{2} \hbar^2)$. This
logarithmic singularity plays a crucial role in the study of phase
separation in Bose-Fermi mixtures  and will be examined in detail
in the following. Note, that the situation is different in 3D
systems, where the density of states is regular, and singularities
appear only in quantities involving derivatives of the density of
states.

As an example we study the dispersion relation for a strong
optical lattice with only nearest neighbor hopping,
\begin{eqnarray}
    \epsilon_{\scriptscriptstyle \rm F}({\bf q})& =& -2 J_{\rm
    \scriptscriptstyle F} \left[ \cos \left(q_{x} \frac{\lambda}{2}\right) +
         \cos \left( q_{y} \frac{\lambda}
         {2} \right) \right]
         \label{fermionicdispersion}
\end{eqnarray}
and  $J_{\rs F}$ the hopping energy.
The density of states is irregular and exhibits a logarithmic van
Hove singularity at $ \epsilon_{0}=0$ arising from saddle points
at  ${\bf k}_0 = (\pi/a,0)$ and ${\bf k}_0 = (0,\pi/a)$,
\begin{equation}
    N(\epsilon) = N_{0} K\left[\sqrt{1-\frac{\epsilon^{2}}
    {16J_{\rs F}^{2}}}\:\right] \sim
    N_{0} \ln\left|\frac{16 J_{\rs F}}{\epsilon} \right|, \label{vanHove}
\end{equation}
with $N_{0}= 1/(2 \pi^{2} J_{\rs F})$, and $K[m]$ the complete
elliptic integral of the first kind \cite{xing91}. The hopping
amplitude $J_{\rs F}$ derives from the exactly known width of the
lowest band in the 1D Mathieu equation \cite{abramowitz}
\begin{equation}
    4J_{\rs F} = \frac{16}{\sqrt{\pi}} \sqrt{E_{\rs F}
    V_{\rs F}}\left(\frac{V_{\rs F}}{E_{\rs F}}\right)^{1/4}
    \exp\left(-2 \sqrt{\frac{V_{\rs F}}{E_{\rs F}}}\right) \label{hoppingJFB} .
\end{equation}

A reliable estimate of the interaction strength is provided by
approximating the Wannier functions $\widetilde{w}({\bf x})$ and
$\widetilde{v}({\bf x})$ by the wave function of the harmonic
oscillator in each well. The oscillator frequency in each well is
given by $\omega_{\rs well}=\sqrt{4 E_{\alpha} V_{\alpha}}/\hbar$
with $\alpha= {\rm B},{\rm F}$, which implies a size $a_{\rs well} =
\sqrt{\hbar/m \omega_{\rs well}}$ for the localized wave function.
The interaction strengths $U_{\rs FB}$ and $U_{\rs B}$, cf.\
(\ref{UB}) and (\ref{UFB}), assume the form
\begin{eqnarray}
    U_{\rs B} &= &\!8 \sqrt{\pi} \:E_{\rs B}\:
    \frac{a_{\rs B}}{\lambda}
    \left(\frac{\hbar \omega_{\rs F\perp}}{2 E_{\rs B}}\right)^{1/2}
    \left( \frac{V_{\rs B}}{E_{\rs B}}\right)^{1/2}, \label{interactionUB}\\
    U_{\rs FB}&= &\!8 \sqrt{\pi}\: \sqrt{E_{\rs B} E_{\rs F}}
    \: \frac{a_{\rs FB}}{\lambda}
    \left(\frac{\hbar \omega_{\rs F \perp}}{2 E_{\rs F}}
    \frac{ \hbar \omega_{\rs B\perp }}{2 E_{\rs
    B}}\right)^{1/4}\!\!\!
   \left(\frac{ V_{\rs F}}{E_{\rs F}} \frac{ V_{\rs B}}{E_{\rs
    B}}\right)^{1/4} \!\!.
\nonumber
\end{eqnarray}
The validity of the derivation of the Hamiltonian
(\ref{tbhamiltonian}) requires that the interaction parameters
$U_{\rs B}$ and $U_{\rs FB}$ are small compared to the energy gap
$\sim \hbar \omega_{\rs well}$ separating the lowest Bloch band
from the next higher.
%
%

\subsection{Effective boson Hamiltonian}

In order to study the stability of the ground state, it is
convenient to derive an effective Hamiltonian for the bosons
alone. This effective Hamiltonian accounts for the fermions via a
modified interaction between the bosons. We start with linear
response theory, where the boson density $n_{\rm
\scriptscriptstyle B}({\bf q})$ drives the fermionic system
$\langle n_{\rm \scriptscriptstyle F}({\bf q})\rangle= U_{\rm
\scriptscriptstyle FB} \chi(T,{\bf q}) n_{\rm \scriptscriptstyle
B}({\bf q})$ with $\chi(T,{\bf q})$ the response function of the
fermions at temperature $T$. This perturbed fermionic density in
turn acts as a drive for the bosons and is accounted for by the
effective interaction between the bosons
\begin{equation}
    H_{\rm \scriptscriptstyle int} = \frac{1}{2N} \!\!\! \sum_{\{{\bf
    k,k',q,q'}\}}\!\!\!\!
    \left[U_{\rm \scriptscriptstyle B}
    + U_{\rm \scriptscriptstyle FB}^{2} \: \chi(T,{\bf q-q'}) \right]
    b^{+}_{{\bf k}} b^{}_{{\bf k}'} b^{+}_{{\bf q}}b^{}_{{\bf q}'}.
    \label{interaction}
\end{equation}
The response function of the fermions is given by the Lindhard
function
\begin{equation}
    \chi(T,{\bf q}) = \int_{K}\frac{d{\bf k}}{v_{0}}
    \frac{f[\epsilon_{\rm \scriptscriptstyle F}({\bf k})]
    -f[\epsilon_{\rm \scriptscriptstyle F}({\bf k+q})] }
    {\epsilon_{\rm \scriptscriptstyle F}({\bf k})-
    \epsilon_{\rm \scriptscriptstyle F}({\bf k+q}) + i \eta}
    \label{lindhardfunction}
\end{equation}
with $v_{0}=(2 \pi/a)^{2}$ the volume of the first Brillouin zone.
The temperature $T$ enters via the Fermi distribution function
$f(\epsilon)=1/\{1+\exp[(\epsilon-\mu_{\rs F})/T]\}$.
We focus on static instabilities of the ground state and neglect
the frequency dependence of the response function. A rigorous
derivation of the effective action including its time dependence
can be achieved within a path integral approach
\cite{buechlerthesis}.

\section{Phase separation \label{phaseseparation}}

\subsection{Instability at $q=0$}

The Lindhard function is always negative and induces an attraction
between the bosons independent of the attractive/repulsive nature
of the original coupling  $U_{\rs FB}$ between the bosons and
fermions. The effective long distance scattering parameter for
${\bf q} \rightarrow 0$ takes the form
\begin{equation}
    U_{\rm \scriptscriptstyle eff}=  U_{\rm \scriptscriptstyle B} +
    U_{\rm \scriptscriptstyle FB}^{2} \: \chi(T,0).
    \label{effectiveinteraction}
\end{equation}
For a fermionic system with a regular density of states, the
Lindhard function  at ${\bf q}=0$ and low temperatures reduces to
$\chi(T \rightarrow 0,0) =-N(\epsilon_{\rs F})$ with $N(\epsilon)$
the fermionic density of states and $\epsilon_{\rs F}$ the Fermi
energy. For fermions on a square lattice in 2D the density of
states is irregular and exhibits a logarithmic van Hove
singularity at $\epsilon_{0}$, see Eq.\ (\ref{vanHove}).
%
%
For a fermionic filling such that the Fermi energy
$\epsilon_{\rs F}$ matches the energy of the van Hove singularity,
i.e. $\epsilon_{\rs F}=\epsilon_{0}$, the Lindhard function
diverges logarithmically for  $T\rightarrow 0$, and its asymptotic
behavior takes the form
\begin{equation}
    \chi(T\rightarrow 0,0)=\int d\epsilon N(\epsilon)
    \frac{\partial f(\epsilon)}{\partial \epsilon}
    =-N_{0}
    \ln\frac{2 c_{1} W_{\rs F}}{T} \label{q0singularity}
\end{equation}
with $c_{1} =2 \exp(C)/\pi\approx 1.13$ a numerical prefactor and
$C\approx 0.577$ the Euler constant.

As a consequence of this logarithmic divergence of the Lindhard
function for $T \rightarrow 0$, the effective scattering parameter
$U_{\rs eff}$ always turns negative at low temperatures. Since a
thermodynamically stable superfluid condensate requires a positive
effective interaction $U_{\rm \scriptscriptstyle eff}>0$
\cite{abrikosovbook}, the system exhibits an instability at the
critical temperature $T_{\rs PS}$ defined via $U_{\rm
\scriptscriptstyle eff}(T_{\rs PS})=0$. Using
Eqs.~(\ref{effectiveinteraction}) and (\ref{q0singularity}), we
find the critical temperature
\begin{equation}
    T_{\rm \scriptscriptstyle PS} = 2 c_{1}
        W_{\rs F} \exp\left[-
        1/\lambda_{\rs FB}\right]\label{TPS}
\end{equation}
with $\lambda_{\rs FB} = (U_{\rs FB}^{2}/U_{\rs B})N_{0}$ the
dimensionless coupling constant. Note, that weak coupling requires
$\lambda_{\rs FB} <1$ and the critical temperature $T_{\rs PS}$ is
well below the transition temperature $T_{\rs KT}$ of the
superfluid condensate. Below the critical temperature $T_{\rs PS}$
the effective interaction $U_{\rs eff}$ turns negative providing a
negative compressibility, and the homogeneous superfluid
condensate becomes unstable. Then, the new ground state with fixed
averaged density $n_{\rs B}$ and $n_{\rs F}$ exhibits phase
separation with areas of increased and decreased local densities
coexisting.

Note, that this transition towards phase separation exhibits two
major differences as compared to the phase separation discussed in
Refs.\ \cite{molmer98,viverit00}. First, the phase separation is
an instability appearing at low temperatures for arbitrary small
coupling $U_{\rs FB}$ between the bosons and fermions. Second, the
increase/decrease in the bosonic density drives the fermionic
density away from a filling factor close to the van Hove
singularity  providing a regular $\chi(T,0)$ which in turn
stabilizes the system.

\subsection{Thomas-Fermi approximation}

In the following, we study the phase separated state within the
Thomas-Fermi theory. Within this theory, we introduce two
densities $n_{\rs B}({\bf x})$ and $n_{\rs F}({\bf x})$ which are
smooth on a scale large compared to the Fermi wave length
$1/k_{\rs F}$ and the bosonic coherence length $\xi = 1/(8\pi n
a_{\rs B})^{1/2}$. Then, the system is in thermodynamic
equilibrium at every position and neglecting the kinetic energy of
the bosons the free energy ${\mathcal F}\left[n_{\rs B}, n_{\rs
F}\right]$ of the system becomes
\begin{eqnarray}
    {\mathcal F}\left[n_{\rs B}, n_{\rs F}\right] &=& \int \frac{d{\bf
    x}}{a^{2}}
    \bigg\{ F_{\rs F}\left[n_{\rs F}({\bf x}) \right] +
    V_{\rs F}^t({\bf x}) n_{\rs F}({\bf x})\label{thomasfermifunctional}\\ & &
    \hspace{-45pt}+ V_{\rs B}^t({\bf x}) n_{\rs B}({\bf x})
    + \frac{1}{2} U_{\rs B} n_{\rs B}({\bf x}) n_{\rs B}({\bf x})+
    U_{\rs FB}n_{\rs B}({\bf x}) n_{\rs F}({\bf x}) \bigg\} . \nonumber
\end{eqnarray}
Here, we include the weak external trapping potential $V_{\rs B}^t$
for the bosons and $V_{\rs F}^t$ for the fermions. Furthermore, the
appearance of the instability requires that the critical
temperature $T_{\rs PS}$ is large compared to the mean level
spacing introduced by the external trapping potentials, i.e.,
$T_{\rs PS}\gg \hbar \omega$ with $\omega$ a characteristic
trapping frequency.
Note, that we
assume $T \ll T_{\rs KT}$ which implies that the influence of
thermally excited bosons is small and can be neglected. The local
free energy $F_{\rs F}[n_{\rs F}]$ of the fermions at temperature
$T$ takes the form ($F=\Omega- \mu n$)
\begin{equation}
  F_{\rs F}[n_{\rs F}]= - T \int d\epsilon
  N(\epsilon)
  \ln \left[1+\exp\left(-\frac{\epsilon-\mu}{T}\right)\right]-\mu
  n_{\rs F} .
\end{equation}
The local chemical potential $\mu({\bf x})$ is determined by the
condition that the fermionic free energy $F_{\rs F}$ is a minimum,
i.e., $\partial_{\mu} F_{\rs F}= 0$, and hence
\begin{equation}
   n_{\rs F}({\bf x})= \int d\epsilon N(\epsilon)
   f\left[\epsilon-\mu({\bf x})\right]. \label{muversusn}
\end{equation}
Furthermore, the total number of particles in the system is fixed
providing the additional constraints
\begin{equation}
  N_{\rs F}= \int d{\bf x} \: n_{\rs F}(x), \hspace{20pt }
   N_{\rs B}= \int d{\bf x}\: n_{\rs B}(x) . \label{constraint}
\end{equation}
The ground state configuration at temperature $T$ is determined by
the minima of the functional ${\mathcal F}[n_{\rs B},n_{\rs F}]$
satisfying the conditions (\ref{constraint}), and hence
\begin{eqnarray}
    \mu\left[n_{\rs F}({\bf x})\right]+ V_{\rs
    F}^t({\bf x}) + U_{\rs FB}n_{\rs B}({\bf x})&= &\mu_{\rs F},
    \label{thomasfermiequation1}\\
    U_{\rs B}n_{\rs B}({\bf x})+ V_{\rs
    B}^t({\bf x}) + U_{\rs FB}n_{\rs F}({\bf x})&= &\mu_{\rs B},
    \label{thomasfermiequation2}
\end{eqnarray}
with $\mu_{\rs F}$ and $\mu_{\rs B}$ the Laplace multipliers
introduced to account for (\ref{constraint}). These parameters act
as global chemical potentials for the system (we have used the
relation $\partial_{n_{\rs F}}F_{\rs F}[n_{\rs F}]=\mu[n_{\rs
F}]$).

In the following, we neglect the trapping potentials $V_{\rs F}^t$
and $V_{\rs B}^t$ and analyze the stability of the state with a
homogeneous fermionic and bosonic density $n_{\rs F}$ and $n_{\rs
B}$. Then, Eqs.~(\ref{thomasfermiequation1}) and
(\ref{thomasfermiequation2}) can be cast into a simpler form. From
Eq.~(\ref{thomasfermiequation2}), we obtain the bosonic density
$n_{\rs B}=(\mu_{\rs B}- U_{\rs FB}n_{\rs F})/U_{\rs B}$ and
inserting this result into Eq.~(\ref{thomasfermiequation1}), we
obtain
\begin{equation}
 \mu[n_{\rs F}]=\mu_{\rs F}-\frac{U_{\rs FB}}{U_{\rs B}}\mu_{\rs
 B}+\frac{U_{\rs FB}^{2}}{U_{\rs B}}n_{\rs F}.
 \label{muF}
\end{equation}
Using (\ref{muversusn}) we arrive at an implicit equation for
$n_{\rs F}$ at fixed global chemical potentials $\mu_{\rs F}$
and $\mu_{\rs B}$,
\begin{equation}
  n_{\rs F}= \int d\epsilon N(\epsilon)\:
  f\left(\epsilon\!-\!\mu_{\rs F}\!
  +\!\frac{U_{\rs FB}}{U_{\rs B}}\mu_{\rs
  B}\!-\!\frac{U_{\rs FB}^{2}}{U_{\rs B}}n_{\rs F}\right).
  \label{integralequation}
\end{equation}
A  solution of (\ref{integralequation}) with fermionic and bosonic
densities $n_{\rs F}$ and $n_{\rs B}$ is a local minimum of the
functional (\ref{thomasfermifunctional}), if the Hessian
\begin{equation}
    {\bf H}=\left(
    \begin{array}{cc}
    \partial_{n_{\rs F}}\mu[n_{\rs F}] &U_{\rs FB}\\
    U_{\rs FB}  &U_{B}
    \end{array}
    \right)
\end{equation}
is positive definite. The derivative $\partial_{n_{\rs
F}}\mu[n_{\rs F}]$ is related to the response function
$\chi(T,{\bf q})$ via the compressibility sum rule and takes the
form
\begin{eqnarray}
  \partial_{n_{\rs F}}\mu[n_{\rs F}]&=&\left\{\int d\epsilon N(\epsilon)
  \left[-\partial_{\epsilon} f(\epsilon-\mu)\right]\right\}^{-1} \\
  &=&
  - \left[\chi(T,0)\right]^{-1}.
\end{eqnarray}
A positive definite Hessian imposes the stability conditions
\begin{eqnarray}
  {\rm Tr} {\bf H}& = &|\chi(T,0)|+U_{B}>0,
  \\ {\rm det} {\bf H}& = &|\chi(T,0)|^{-1} \:U_{B}-U_{\rs
  FB}^{2}>0. \label{stabilityconditions}
\end{eqnarray}
The first condition is always satisfied for a repulsive
interaction $U_{\rs B}>0$ between the bosons. For a regular
density of states $N(\epsilon)$, the second stability condition at
zero temperature $T=0$ becomes
\begin{equation}
  \frac{U_{\rs FB}^{2}}{U_{\rs B}}\: N(\epsilon_{F})=1
\end{equation}
and coincides with the condition for phase separation in Refs.\
\onlinecite{molmer98,viverit00}; $\epsilon_{\rs F}$ denotes the Fermi
energy.

However, for the system considered here with 2D fermions on a
square lattice the density of states exhibits a van Hove
singularity. A setup with a fermionic density $\overline{n}_{\rs
F}$ matching the van Hove singularity at $\epsilon_0$ and
arbitrary bosonic density $\overline{n}_{\rs B}$ then is of
special interest. These densities have to satisfy the conditions
(\ref{thomasfermiequation1}) and (\ref{thomasfermiequation2}),
from which we find the appropriate global chemical potentials
$\overline{\mu}_{\rs B}=U_{\rs FB}\overline{n}_{\rs F} +U_{\rs
B}\overline{n}_{\rs B}$ and, using $\mu[\overline{n}_{\rs F}] =
\epsilon_0$, $\overline{\mu}_{\rs F}=U_{\rs FB}\overline{n}_{\rs
B} +\epsilon_{0}$. Fixing the global chemical potentials
$\overline{\mu}_{\rs F}$ and $\overline{\mu}_{\rs B}$ rather than
the densities $\overline{n}_{\rs F}$ and $\overline{n}_{\rs B}$,
we can allow for the appearance of new solutions, i.e., phase
separation. The new fermion density $n_{\rs F}$ then has to
satisfy the condition (\ref{integralequation}) at fixed values
$\overline{\mu}_{\rs B}$ and $\overline{\mu}_{\rs F}$ of the
chemical potentials,
\begin{equation}
  n_{\rs F}= \int d\epsilon N(\epsilon)\:
  f\left(\epsilon-\epsilon_0-\frac{U_{\rs FB}^{2}}{U_{\rs B}}
  (n_{\rs F}-\overline{n}_{\rs F})\right).
\label{isochore}
\end{equation}
The self-consistency equation (\ref{isochore}) has
the obvious solution $n_{\rs F}=\overline{n}_{\rs F}$. This
solution turns unstable at the point where the second stability
condition (\ref{stabilityconditions}) is violated: defining
the deviation $\delta \overline{n}_{\rs F}= n_{\rs F}-
\overline{n}_{\rs F}$ and expanding (\ref{isochore})
around $\overline{n}_{\rs F}$ we find that
\begin{equation}
   \alpha(T)\; \delta \overline{n}_{\rs F}
   + \beta(T) \big(\delta \overline{n}_{\rs F}\big)^{2}
   + \gamma(T) \; \big(\delta \overline{n}_{\rs F}\big)^{3} = 0,
   \label{expansion}
\end{equation}
where $\alpha(T)=[1+ (U^{2}_{\rs FB}/U_{\rs B})\int d\epsilon
N(\epsilon+\epsilon_{0}) \partial_{\epsilon} f(\epsilon)] \approx
\alpha_{0} (T-T_{\rs PS})$, while $\beta(T)$ and $\gamma(T)>0$
depend only weakly on temperature. For a system with particle-hole
symmetry the quadratic term vanishes, i.e., $\beta(T) = 0$; this
is the case for the situation discussed above with the dispersion
relation (\ref{fermionicdispersion}) and a half-filled band with
Fermi energy $\epsilon_{\rs F}=0$. The sign change in $\alpha(T)$
at $T_{\rs PS}$ matches up with the violation of
(\ref{stabilityconditions}) and the expression for the transition
temperature $T_{\rs PS}$ agrees with the one derived previously in
(\ref{TPS}). At high temperatures $T>T_{\rs PS}$, $\alpha>0$ and
Eq.~(\ref{isochore}) has only the trivial solution
$\overline{n}_{\rs F}=0$. For low temperatures $T<T_{\rs PS}$,
$\alpha<0$ and two new solutions $\overline{n}_{\rs F} \pm \delta
\overline{n}_{\rs F}$ appear with
\begin{equation}
   \delta \overline{n}_{\rs F}= \sqrt{\alpha_{0}/\gamma}\: (T_{\rs
   PS}- T)^{1/2},
   \label{dn_Tc}
\end{equation}
where we assume $\beta = 0$ and $T$ close to $T_{\rs PS}$.

At zero temperature and for $\lambda_{\rs FB} \ll 1$ we find
the density change
\begin{eqnarray}
 \delta \overline{n}_{\rs F} &= &\frac{U_{\rs FB}^{2}}{U_{\rs
 B}}\int_{0}^{\delta n_{\rs F}} dn N\left(\epsilon_{0}+\frac{U_{\rs
 FB}^{2}}{U_{\rs B}} n\right) \label{dn_0}\\
 && \hspace{50pt}\sim \delta \overline{n}_{\rs F}
 \lambda_{\rs FB } \ln\left(\frac{16 c_{2} J_{\rs F} U_{\rs
 B}}{U_{\rs FB}^{2} |\delta \overline{n}_{\rs F}|} \right)
 \nonumber
\end{eqnarray}
with $c_{2}=\exp(1)$. Solving for $\delta \overline{n}_{\rs F}$
provides us with the following shifts in the fermionic and bosonic
densities
\begin{eqnarray}
  n_{\rs F}-\overline{n}_{\rs F}&=&\delta \overline{n}_{\rs F} =
  \pm 16  c_{2} \frac{U_{\rs B} J_{\rs F}}
  {U_{\rs FB}^{2}} \exp\left(-\frac{1}{\lambda_{\rs FB}}\right),  \label{solutions}\\
  n_{\rs B}-\overline{n}_{\rs B}& = &-\frac{U_{\rs FB}}{U_{\rs B}}
  \delta \overline{n}_{\rs F} =\mp 16  c_{2}\frac{J_{\rs F}}
  {U_{\rs FB}} \exp\left(-\frac{1}{\lambda_{\rs FB}}\right).
\nonumber
\end{eqnarray}
The relation between the $T=0$ density shift $\delta
\overline{n}_{\rs F}$ and the critical temperature $T_{\rs PS}$
takes the form
\begin{equation}
   \frac{U_{\rs FB}^{2}}{U_{\rs B}} \frac{\delta \overline{n}_{\rs F}}{T_{\rs
   PS}}= c_{2}/c_{1} \approx 2.40 .
\end{equation}
Inserting the solutions (\ref{solutions}) into the free energy
(\ref{thomasfermifunctional}) provides us with the energy shift
per unit cell at zero temperature
\begin{equation}
\frac{\Delta {\mathcal F}}{N} = - \frac{U_{\rs FB}^{2}}{2 U_{\rs
B}} \left(\delta \overline{n}_{\rs F}\right)^{2} .
\label{phaseseparationcondensationenergy}
\end{equation}

\subsection{Phase diagram}

The solution discussed above allows us to find the phase diagram
of the system. At low temperature $T<T_{\rs PS}$, we distinguish
between a low-density `gas' phase with $n_{\rs F}\leq
\overline{n}_{\rs F}-\delta \overline{n}_{\rs F}$, and a
high-density `fluid' phase with $n_{\rs F}\geq \overline{n}_{\rs
F}+\delta \overline{n}_{\rs F}$. The bosonic density derives from
$n_{\rs B}=(\mu_{\rs B}-U_{\rs FB} n_{\rs F})/U_{\rs B}$. For a
repulsive interaction $U_{\rs FB}>0$ the fermionic low-density
phase partners up with a high bosonic density, while for attractive
interaction $U_{\rs FB}<0$ the fermionic low-density phase implies
a low bosonic density.
\begin{figure}[hbtp]
\begin{center}
\includegraphics[scale=0.52]{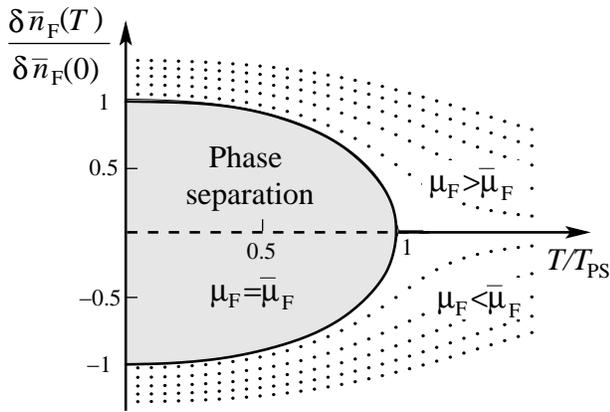}
  \caption[Phase diagram of the transition towards phase
  separation]{$n_{\rs F}$-$T$ phase diagram. The grey region
  denotes the phase separated regime,  while the solid line
  denotes the line of fixed chemical potential $\overline{\mu}_{\rs
  F}$. The dotted lines describe fixed chemical potentials
  $(\mu_{\rs F}-\overline{\mu}_{\rs F})/T_{\rs PS}=\pm
  0.02,0.06,0.01,0.14,0.18$. The lines derive from
  Eq.~(\ref{integralequation}) with $\lambda=0.2$ and the
  dispersion relation (\ref{fermionicdispersion}).
  \label{PSMeanfield}}
  \end{center}
\end{figure}

The phase transition is of the standard liquid-gas transition
type, see Fig.~\ref{PSMeanfield}. At temperatures below
critical, $T<T_{\rs PS}$, the low-density `gas' phase
is separated by a first-order phase transition from the
high-density `liquid' phase. The transition is driven by
the chemical potential and takes place at the critical value
\begin{equation}
   \overline{\mu}_{\rs F}
   = \frac{U_{\rs FB}}{U_{\rs B}} \overline{\mu}_{\rs B}-
   \frac{ U_{\rs FB}^{2}}{2 U_{\rs B}},
\end{equation}
cf.\ Eq.\ (\ref{muF}), where we have used the results $\epsilon_0
= 0$ and $\overline{n}_{\rs F} = 1/2$ valid for the dispersion
(\ref{fermionicdispersion}). A fixed averaged fermionic density
between $\overline{n}_{\rs F}-\delta \overline{n}_{\rs F}<n_{\rs
F}<\overline{n}_{\rs F}+\delta \overline{n}_{\rs F}$ is only
realized via coexistence of the low density `gas' phase and the
high density `liquid' phase. The first order transition terminates
in a critical endpoint at the temperature $T_{\rs PS}$ and density
$n_{\rs F}=\overline{n}_{\rs F}$. Varying the temperature across
the critical value $T_{\rs PS}$ along the isochore $n_{\rs
F}=\overline{n}_{\rs F}$, we obtain a second-order phase
transition between the homogeneous phase and the phase separated
state, see Fig.~\ref{PSMeanfield}. This transition appears for
arbitrary weak coupling $U_{\rs FB}$ due to the enhanced fermionic
density of states for 2D fermions.

%
\begin{figure}[hbtp]
\begin{center}
\includegraphics[scale=0.35]{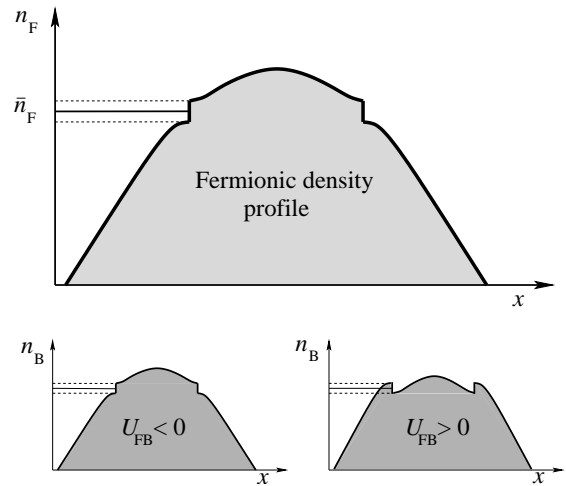}
  \caption{Sketch of the fermionic and bosonic density profiles
  in the weak coupling limit $\lambda_{\rs FB}\ll 1$.
  (a) The fermionic density exhibits a
  jump in the density by $2 \delta \overline{n}_{F}$ as the first
  order transition line is crossed. At the same position in space,
  also the bosonic density exhibits a jump. For attractive
  interaction $U_{\rs FB}<0$ between the bosons and the fermions
  the bosonic density increases, see (b),
  while for repulsive interaction $U_{\rs FB}>0$
  the bosonic density decreases at the jump (c).
  \label{densityProfile}}
  \end{center}
\end{figure}

\section{Finite trapping potential and conclusions \label{conclusions}}

We have shown that optical lattices strongly modify the behavior
of Bose-Fermi mixtures via the appearance of van Hove
singularities. The effect is most pronounced in two-dimensions,
where the density of states diverges logarithmically, see
Eq.~(\ref{densityofstates}). Then, for a fermionic chemical
potential matching the position of the van Hove singularity, i.e.,
$\mu= \epsilon_{0}$, the mixture of bosons and fermions undergoes
a second order phase transition at the critical temperature
$T_{\rs PS}$. In analogy to the standard `liquid--gas' gas
transition, a fermionic density $n_{\rs F}$ with
$\overline{n}_{\rs F}-\delta \overline{n}_{\rs F}<n_{\rs
F}<\overline{n}_{\rs F}+\delta \overline{n}_{\rs F}$ is unstable
and inaccessible in an experiment for temperatures $T<T_{\rs PS}$.
The fermionic density profile in a harmonic trap can  be derived
from Eqs.~(\ref{thomasfermiequation1}) and
(\ref{thomasfermiequation2}), and its behavior at zero temperature
and weak coupling is sketched in Fig.~\ref{densityProfile}. The
weak coupling $\lambda_{\rs FB}\ll 1$ between the fermions and
bosons makes sure that the density profile is only slightly
modified compared to the noninteracting limit. As the local
chemical potential $\mu(x)$ matches up with the van Hove
singularity $\epsilon_{0}$, the fermionic density exhibits a jump
by $2 \delta \overline{n}_{\rs F}$ related to the crossing of the
first-order transition line. This behavior does not depend on the
sign of the interaction $U_{\rs FB}$ between the bosons and
fermions.
%
%
%
In turn, the bosonic density profile strongly differs for
attractive and repulsive $U_{\rs FB}$. At the same position as the
fermionic density profile exhibits a jump in the density, the
bosonic density profile also exhibits a jump by $- 2 U_{\rs FB}
\delta \overline{n}_{\rs F}/U_{\rs  B}$.  The bosonic density
increases similarly to the fermionic density profile for an
attractive interaction $U_{\rs FB}<0$ between the bosons and
fermions, while it is decreased for a repulsive interaction
$U_{\rs FB}>0$, see Fig.~\ref{densityProfile}. The different
behavior of bosons and fermions is a consequence of the asymmetric
role played by the constituents: the transition is driven by the
van Hove singularity in the fermionic density of states and is
independent on the bosonic density. Note that the situation is
quite different in the strong coupling limit $\lambda_{\rs FB}
\approx 1$. Then the van Hove singularity in the density of states
plays a minor role and the nature of the phase separated state is
dominated by the character of the interaction and the trapping
potential, a situation previously discussed in
Refs.~\onlinecite{molmer98,nygaard99,modugno02,amoruso98,roth02,viverit00}.

We have shown that the presence of van Hove singularities
in the fermionic density of states produces new and
interesting features in the context of phase separation in
a 2D Bose-Fermi mixture. At the same time, these van Hove
singularities also enhance the instabilities driving
other quantum phases competing with phase separation.
E.g., the instability towards BCS-superconductivity
\cite{hirsch86} or the recent proposal for a supersolid phase
\cite{buechler03.2} are both driven by Fermi-surface nesting,
usually involving a $\ln(\epsilon_{\rs F}/T)$ divergence which
is enhanced to a $[\ln(\epsilon_{\rs F}/T)]^{2}$ singularity
in the presence of a van Hove singularity. 2D Bose-Fermi
mixtures then are promising candidates for the observation
of such new quantum phases, provided they successfully compete
against the tendency towards phase separation lurking at
$T_{\rs PS}$.


\begin{thebibliography}{22}
\expandafter\ifx\csname natexlab\endcsname\relax\def\natexlab#1{#1}\fi
\expandafter\ifx\csname bibnamefont\endcsname\relax
  \def\bibnamefont#1{#1}\fi
\expandafter\ifx\csname bibfnamefont\endcsname\relax
  \def\bibfnamefont#1{#1}\fi
\expandafter\ifx\csname citenamefont\endcsname\relax
  \def\citenamefont#1{#1}\fi
\expandafter\ifx\csname url\endcsname\relax
  \def\url#1{\texttt{#1}}\fi
\expandafter\ifx\csname urlprefix\endcsname\relax\def\urlprefix{URL }\fi
\providecommand{\bibinfo}[2]{#2}
\providecommand{\eprint}[2][]{\url{#2}}

\bibitem[{\citenamefont{Truscott et~al.}(2001)\citenamefont{Truscott, Strecker,
  McAlexander, Partridge, and Hulet}}]{truscott01}
\bibinfo{author}{\bibfnamefont{A.~G.} \bibnamefont{Truscott}},
  \bibinfo{author}{\bibfnamefont{K.~E.} \bibnamefont{Strecker}},
  \bibinfo{author}{\bibfnamefont{W.~I.} \bibnamefont{McAlexander}},
  \bibinfo{author}{\bibfnamefont{G.~B.} \bibnamefont{Partridge}},
  \bibnamefont{and} \bibinfo{author}{\bibfnamefont{R.~G.} \bibnamefont{Hulet}},
  \bibinfo{journal}{Science} \textbf{\bibinfo{volume}{291}},
  \bibinfo{pages}{2570} (\bibinfo{year}{2001}).

\bibitem[{\citenamefont{Schreck et~al.}(2001)\citenamefont{Schreck, Khaykovich,
  Corwin, Ferrari, Bourdel, Cubizolles, and Salomon}}]{schreck01}
\bibinfo{author}{\bibfnamefont{F.}~\bibnamefont{Schreck}},
  \bibinfo{author}{\bibfnamefont{L.}~\bibnamefont{Khaykovich}},
  \bibinfo{author}{\bibfnamefont{K.~L.} \bibnamefont{Corwin}},
  \bibinfo{author}{\bibfnamefont{G.}~\bibnamefont{Ferrari}},
  \bibinfo{author}{\bibfnamefont{T.}~\bibnamefont{Bourdel}},
  \bibinfo{author}{\bibfnamefont{J.}~\bibnamefont{Cubizolles}},
  \bibnamefont{and} \bibinfo{author}{\bibfnamefont{C.}~\bibnamefont{Salomon}},
  \bibinfo{journal}{Phys.\ Rev.\ Lett.} \textbf{\bibinfo{volume}{87}},
  \bibinfo{pages}{080403} (\bibinfo{year}{2001}).

\bibitem[{\citenamefont{Hadzibabic et~al.}(2002)\citenamefont{Hadzibabic, Stan,
  Dieckmann, Gupta, Zwierlein, G{\"o}rlitz, and Ketterle}}]{hadzibabic02}
\bibinfo{author}{\bibfnamefont{Z.}~\bibnamefont{Hadzibabic}},
  \bibinfo{author}{\bibfnamefont{C.~A.} \bibnamefont{Stan}},
  \bibinfo{author}{\bibfnamefont{K.}~\bibnamefont{Dieckmann}},
  \bibinfo{author}{\bibfnamefont{S.}~\bibnamefont{Gupta}},
  \bibinfo{author}{\bibfnamefont{M.~W.} \bibnamefont{Zwierlein}},
  \bibinfo{author}{\bibfnamefont{A.}~\bibnamefont{G{\"o}rlitz}},
  \bibnamefont{and} \bibinfo{author}{\bibfnamefont{W.}~\bibnamefont{Ketterle}},
  \bibinfo{journal}{Phys.\ Rev.\ Lett.} \textbf{\bibinfo{volume}{88}},
  \bibinfo{pages}{160401} (\bibinfo{year}{2002}).

\bibitem[{\citenamefont{Roati et~al.}(2002)\citenamefont{Roati, Riboli,
  Modugno, and Inguscio}}]{roati02}
\bibinfo{author}{\bibfnamefont{G.}~\bibnamefont{Roati}},
  \bibinfo{author}{\bibfnamefont{F.}~\bibnamefont{Riboli}},
  \bibinfo{author}{\bibfnamefont{G.}~\bibnamefont{Modugno}}, \bibnamefont{and}
  \bibinfo{author}{\bibfnamefont{M.}~\bibnamefont{Inguscio}},
  \bibinfo{journal}{Phys.\ Rev.\ Lett.} \textbf{\bibinfo{volume}{89}},
  \bibinfo{pages}{150403} (\bibinfo{year}{2002}).

\bibitem[{\citenamefont{Houbiers et~al.}(1997)\citenamefont{Houbiers, Ferwerda,
  and Stoof}}]{houbiers97}
\bibinfo{author}{\bibfnamefont{M.}~\bibnamefont{Houbiers}},
  \bibinfo{author}{\bibfnamefont{R.}~\bibnamefont{Ferwerda}}, \bibnamefont{and}
  \bibinfo{author}{\bibfnamefont{H.~T.~C.} \bibnamefont{Stoof}},
  \bibinfo{journal}{Phys.\ Rev.\ A} \textbf{\bibinfo{volume}{56}},
  \bibinfo{pages}{4864} (\bibinfo{year}{1997}).

\bibitem[{\citenamefont{Heiselberg et~al.}(2000)\citenamefont{Heiselberg,
  Pethick, Smith, and Viverit}}]{heiselberg00}
\bibinfo{author}{\bibfnamefont{H.}~\bibnamefont{Heiselberg}},
  \bibinfo{author}{\bibfnamefont{C.~J.} \bibnamefont{Pethick}},
  \bibinfo{author}{\bibfnamefont{H.}~\bibnamefont{Smith}}, \bibnamefont{and}
  \bibinfo{author}{\bibfnamefont{L.}~\bibnamefont{Viverit}},
  \bibinfo{journal}{Phys.\ Rev.\ Lett.} \textbf{\bibinfo{volume}{85}},
  \bibinfo{pages}{2418} (\bibinfo{year}{2000}).

\bibitem[{\citenamefont{Hofstetter et~al.}(2002)\citenamefont{Hofstetter,
  Cirac, Zoller, Demler, and Lukin}}]{hofstetter02}
\bibinfo{author}{\bibfnamefont{W.}~\bibnamefont{Hofstetter}},
  \bibinfo{author}{\bibfnamefont{J.~I.} \bibnamefont{Cirac}},
  \bibinfo{author}{\bibfnamefont{P.}~\bibnamefont{Zoller}},
  \bibinfo{author}{\bibfnamefont{E.}~\bibnamefont{Demler}}, \bibnamefont{and}
  \bibinfo{author}{\bibfnamefont{M.~D.} \bibnamefont{Lukin}},
  \bibinfo{journal}{Phys.\ Rev.\ Lett.} \textbf{\bibinfo{volume}{89}},
  \bibinfo{pages}{220407} (\bibinfo{year}{2002}).

\bibitem[{\citenamefont{Viverit}(2002)}]{viverit02}
\bibinfo{author}{\bibfnamefont{L.}~\bibnamefont{Viverit}},
  \bibinfo{journal}{Phys.\ Rev.\ A} \textbf{\bibinfo{volume}{66}},
  \bibinfo{pages}{023605} (\bibinfo{year}{2002}).

\bibitem[{\citenamefont{B{\"u}chler and Blatter}(2003)}]{buechler03.2}
\bibinfo{author}{\bibfnamefont{H.~P.} \bibnamefont{B{\"u}chler}}
  \bibnamefont{and} \bibinfo{author}{\bibfnamefont{G.}~\bibnamefont{Blatter}},
  \bibinfo{journal}{Phys.\ Rev.\ Lett.} \textbf{\bibinfo{volume}{91}},
  \bibinfo{pages}{130404} (\bibinfo{year}{2003}).

\bibitem[{\citenamefont{Lewenstein et~al.}(2003)\citenamefont{Lewenstein,
  Santos, Baranov, and Fehrmann}}]{lewenstein03}
\bibinfo{author}{\bibfnamefont{M.}~\bibnamefont{Lewenstein}},
  \bibinfo{author}{\bibfnamefont{L.}~\bibnamefont{Santos}},
  \bibinfo{author}{\bibfnamefont{M.~A.} \bibnamefont{Baranov}},
  \bibnamefont{and} \bibinfo{author}{\bibfnamefont{H.}~\bibnamefont{Fehrmann}},
  \bibinfo{journal}{cond-mat/0307635}  (\bibinfo{year}{2003}).

\bibitem[{\citenamefont{M{\o}lmer}(1998)}]{molmer98}
\bibinfo{author}{\bibfnamefont{K.}~\bibnamefont{M{\o}lmer}},
  \bibinfo{journal}{Phys.\ Rev.\ Lett.} \textbf{\bibinfo{volume}{80}},
  \bibinfo{pages}{1804} (\bibinfo{year}{1998}).

\bibitem[{\citenamefont{Amoruso et~al.}(1998)\citenamefont{Amoruso, Minguzzi,
  Stringari, Tosi, and Vichi}}]{amoruso98}
\bibinfo{author}{\bibfnamefont{M.}~\bibnamefont{Amoruso}},
  \bibinfo{author}{\bibfnamefont{A.}~\bibnamefont{Minguzzi}},
  \bibinfo{author}{\bibfnamefont{S.}~\bibnamefont{Stringari}},
  \bibinfo{author}{\bibfnamefont{M.~P.} \bibnamefont{Tosi}}, \bibnamefont{and}
  \bibinfo{author}{\bibfnamefont{L.}~\bibnamefont{Vichi}},
  \bibinfo{journal}{Eur. Phys. J. D} \textbf{\bibinfo{volume}{4}},
  \bibinfo{pages}{261} (\bibinfo{year}{1998}).

\bibitem[{\citenamefont{Nygaard and M{\o}lmer}(1999)}]{nygaard99}
\bibinfo{author}{\bibfnamefont{N.}~\bibnamefont{Nygaard}} \bibnamefont{and}
  \bibinfo{author}{\bibfnamefont{K.}~\bibnamefont{M{\o}lmer}},
  \bibinfo{journal}{Phys.\ Rev.\ A} \textbf{\bibinfo{volume}{59}},
  \bibinfo{pages}{2974} (\bibinfo{year}{1999}).

\bibitem[{\citenamefont{Modugno et~al.}(2002)\citenamefont{Modugno, Ferrari,
  Roati, Brecha, Simoni, and Inguscio}}]{modugno02}
\bibinfo{author}{\bibfnamefont{G.}~\bibnamefont{Modugno}},
  \bibinfo{author}{\bibfnamefont{G.}~\bibnamefont{Ferrari}},
  \bibinfo{author}{\bibfnamefont{G.}~\bibnamefont{Roati}},
  \bibinfo{author}{\bibfnamefont{R.~J.} \bibnamefont{Brecha}},
  \bibinfo{author}{\bibfnamefont{A.}~\bibnamefont{Simoni}}, \bibnamefont{and}
  \bibinfo{author}{\bibfnamefont{M.}~\bibnamefont{Inguscio}},
  \bibinfo{journal}{Science} \textbf{\bibinfo{volume}{297}},
  \bibinfo{pages}{1320} (\bibinfo{year}{2002}).

\bibitem[{\citenamefont{Viverit et~al.}(2000)\citenamefont{Viverit, Pethick,
  and Smith}}]{viverit00}
\bibinfo{author}{\bibfnamefont{L.}~\bibnamefont{Viverit}},
  \bibinfo{author}{\bibfnamefont{C.~J.} \bibnamefont{Pethick}},
  \bibnamefont{and} \bibinfo{author}{\bibfnamefont{H.}~\bibnamefont{Smith}},
  \bibinfo{journal}{Phys.\ Rev.\ A} \textbf{\bibinfo{volume}{61}},
  \bibinfo{pages}{053605} (\bibinfo{year}{2000}).

\bibitem[{\citenamefont{Roth and Feldmeier}(2002)}]{roth02}
\bibinfo{author}{\bibfnamefont{R.}~\bibnamefont{Roth}} \bibnamefont{and}
  \bibinfo{author}{\bibfnamefont{H.}~\bibnamefont{Feldmeier}},
  \bibinfo{journal}{Phys.\ Rev.\ A} \textbf{\bibinfo{volume}{65}},
  \bibinfo{pages}{021603} (\bibinfo{year}{2002}).

\bibitem[{\citenamefont{Hirsch and Scalapino}(1986)}]{hirsch86}
\bibinfo{author}{\bibfnamefont{J.~E.} \bibnamefont{Hirsch}} \bibnamefont{and}
  \bibinfo{author}{\bibfnamefont{D.~J.} \bibnamefont{Scalapino}},
  \bibinfo{journal}{Phys.\ Rev.\ Lett.} \textbf{\bibinfo{volume}{56}},
  \bibinfo{pages}{2732} (\bibinfo{year}{1986}).

\bibitem[{\citenamefont{Jaksch et~al.}(1998)\citenamefont{Jaksch, Bruder,
  Cirac, Gardiner, and Zoller}}]{jaksch98}
\bibinfo{author}{\bibfnamefont{D.}~\bibnamefont{Jaksch}},
  \bibinfo{author}{\bibfnamefont{C.}~\bibnamefont{Bruder}},
  \bibinfo{author}{\bibfnamefont{J.~I.} \bibnamefont{Cirac}},
  \bibinfo{author}{\bibfnamefont{C.~W.} \bibnamefont{Gardiner}},
  \bibnamefont{and} \bibinfo{author}{\bibfnamefont{P.}~\bibnamefont{Zoller}},
  \bibinfo{journal}{Phys.\ Rev.\ Lett.} \textbf{\bibinfo{volume}{81}},
  \bibinfo{pages}{3108} (\bibinfo{year}{1998}).

\bibitem[{\citenamefont{Xing et~al.}(1991)\citenamefont{Xing, Liu, and
  Gong}}]{xing91}
\bibinfo{author}{\bibfnamefont{D.~Y.} \bibnamefont{Xing}},
  \bibinfo{author}{\bibfnamefont{M.}~\bibnamefont{Liu}}, \bibnamefont{and}
  \bibinfo{author}{\bibfnamefont{C.~D.} \bibnamefont{Gong}},
  \bibinfo{journal}{Phys.\ Rev.\ B} \textbf{\bibinfo{volume}{44}},
  \bibinfo{pages}{12525} (\bibinfo{year}{1991}).

\bibitem[{\citenamefont{Abramowitz and Stegun}(1972)}]{abramowitz}
\bibinfo{author}{\bibfnamefont{M.}~\bibnamefont{Abramowitz}} \bibnamefont{and}
  \bibinfo{author}{\bibfnamefont{I.~A.} \bibnamefont{Stegun}},
  \emph{\bibinfo{title}{Handbook of Mahtematical Functions}}
  (\bibinfo{publisher}{Dover Publications}, \bibinfo{address}{New York},
  \bibinfo{year}{1972}).

\bibitem[{\citenamefont{B{\"u}chler}(2003)}]{buechlerthesis}
\bibinfo{author}{\bibfnamefont{H.~P.} \bibnamefont{B{\"u}chler}}, Ph.D. thesis,
  \bibinfo{school}{Swiss Federal Institute of Technology Z{\"u}rich}
  (\bibinfo{year}{2003}).

\bibitem[{\citenamefont{Abrikosov et~al.}(1963)\citenamefont{Abrikosov, Gorkov,
  and Dzyaloshinski}}]{abrikosovbook}
\bibinfo{author}{\bibfnamefont{A.~A.} \bibnamefont{Abrikosov}},
  \bibinfo{author}{\bibfnamefont{L.~P.} \bibnamefont{Gorkov}},
  \bibnamefont{and} \bibinfo{author}{\bibfnamefont{I.~E.}
  \bibnamefont{Dzyaloshinski}}, \emph{\bibinfo{title}{Methods of Quantum Field
  Theory in Statistical Physics}} (\bibinfo{publisher}{Dover Publications},
  \bibinfo{address}{180 Varick Street, New-York 10014}, \bibinfo{year}{1963}).

\end{thebibliography}

\end{document}